\documentclass[aps,prm,twocolumn,twoside,superscriptaddress]{revtex4}

\usepackage[version=3]{mhchem} 
\usepackage{graphicx}
\usepackage{amsmath}
\usepackage{float}
\usepackage{amssymb}
\usepackage{placeins}
\usepackage{array}
\usepackage{subfig}
\usepackage{threeparttable}
\usepackage{color}
\usepackage[symbol]{footmisc}
\usepackage{physics}
\usepackage{makecell}
\usepackage{mhchem}
\usepackage{siunitx}
\usepackage{xr}
\usepackage{xstring}
\usepackage{pdfpages}

\def\co{Co$_\text{3}$O$_\text{4}$}

\begin{document}

    \author{Tyler J. Smart}
    \affiliation{Department of Physics, University of California Santa Cruz, Santa Cruz, CA, 95064, USA}
    \affiliation{Quantum Simulations Group, Lawrence Livermore National Laboratory, Livermore CA, 94551, USA}
    \author{Tuan Anh Pham}
    \affiliation{Quantum Simulations Group, Lawrence Livermore National Laboratory, Livermore CA, 94551, USA}
    \author{Yuan Ping \footnote[1]{yuanping@ucsc.edu}}
    \affiliation{Department of Chemistry and Biochemistry, University of California Santa Cruz, Santa Cruz, CA, 95064, USA}
    \author{Tadashi Ogitsu \footnote[2]{ogitsu1@llnl.gov}}
    \affiliation{Quantum Simulations Group, Lawrence Livermore National Laboratory, Livermore CA, 94551, USA}

    \title[]{
    Optical Absorption Induced by Small Polaron Formation in Transition \\ Metal Oxides - The Case of \co\
    }
    
\begin{abstract}
Small polarons (SPs) are known to exist in most important transition metal oxides (TMOs); however, the nature of small polaron formation remains enigmatic, and a fundamental understanding of how SPs impact the intrinsic electronic structure and optical properties of these materials is largely lacking. In this work, we employ first-principles calculations to investigate SP formation in \co, a highly promising material for a wide range of emerging energy applications, and we resolve the conflicting findings that have been reported on the electronic structure of the system. We confirm that the intrinsic band gap of \co\ is 1.6~eV, and we show that the formation of hole small polarons significantly influences the optical absorption spectra, leading to a 0.8~eV transition that is often misinterpreted as the band edge that defines the fundamental gap. In addition, we discuss how uniaxial strain can be utilized to probe the Jahn-Teller distortion of SP states and in turn, effect their optical transitions. Beyond \co, our study suggests a general roadmap for establishing a first-principles computational approach that can simultaneously achieve an accurate description of SP states, electronic band structure and optical transitions of polaronic magnetic oxides.
\end{abstract}  

\maketitle


Polarons, conduction electrons or holes with self-induced lattice polarization, are known to exist in most transition metal oxides (TMO) and deeply affect their optical and carrier transport properties \cite{reticcioli2019small}. In these materials, much of the interest has been related to the role of small polarons (SPs) that form when the induced lattice polarization is localized in a volume on the order of the unit cell. In particular, for many important TMOs, including \ce{Fe2O3}~\cite{sivula2011solar,ling2011sn,smart2017}, \ce{NiO}~\cite{gong2014nanoscale,hu2014efficient}, \ce{Co3O4}~\cite{wang2018,aijaz2016}, \ce{MnO}~\cite{jin2015partially}, \ce{BiVO4}~\cite{wu2018,zhang2018,seo2018,kim2015}, \ce{CuO}~\cite{cardiel2017electrochemical,smart2018-2}, it has been found that the formation of SPs is responsible for the low carrier mobility and conductivity, which hinders their practical application as electrochemical catalysts and photoelectrochemical (PEC) electrodes \cite{lee2019progress,tachibana2012artificial,roger2017earth,yan2016review, liao2013new}. It is also well-established that the transport of SPs in these TMOs can be characterized through the thermally activated hopping conduction mechanism and a logarithmic temperature dependence of the materials carrier mobility \cite{mott1968conduction}.

Unlike the distinct signature of SPs on the carrier conduction discussed above, the effect of polarons on electronic structure and optical transitions in TMOs is rather complex and difficult to elucidate. For example, in several TMOs, such as WO$_3$~\cite{gerosa2018role,Ping2013}, TiO$_2$~\cite{TiO22018,Moser2013,Kang2010} and SrTiO$_3$~\cite{verdi2017origin,Wang2016}, the presence of large polarons that are delocalized over several unit cells may lead to a strong band gap renormalization through electron-phonon coupling. By contrast, the formation of SPs may introduce isolated gap states away from the band edges due to their spatially localized nature, which could be easily misinterpreted as band edges that define the fundamental band gap. A prime example is \ce{Fe2O3}, where recent time-resolved spectroscopy experiments have shown that its mid-gap states are indeed associated with optically active polarons, which in turn lead to transition energies that are significantly lower than the fundamental band gap \cite{carneiro2017excitation,biswas2018ultrafast}. This conclusion is also consistent with Lohaus \textit{et al.} \cite{lohaus2018limitation}, where the authors showed that while the band gap of \ce{Fe2O3} is 2.2~eV, an effective gap of 1.75~eV is observed due to the formation of small polarons. 
Nonetheless, distinguishing mid-gap states due to SP formation from other sources such as defect-bound states \cite{huda2010electronic,neufeld2015platinum,sanson2015polaronic} and surface states \cite{yatom2015toward} is still not well understood in literature. Despite extensive experiments on these TMOs, theoretical studies of SP effects on electronic structure and optical properties of TMOs are limited. More importantly, challenges remain in first-principles methods that accurately describe polarons and electronic structure of TMOs.

\begin{figure}[t]
\begin{center}
\includegraphics[keepaspectratio=true,width=6cm]{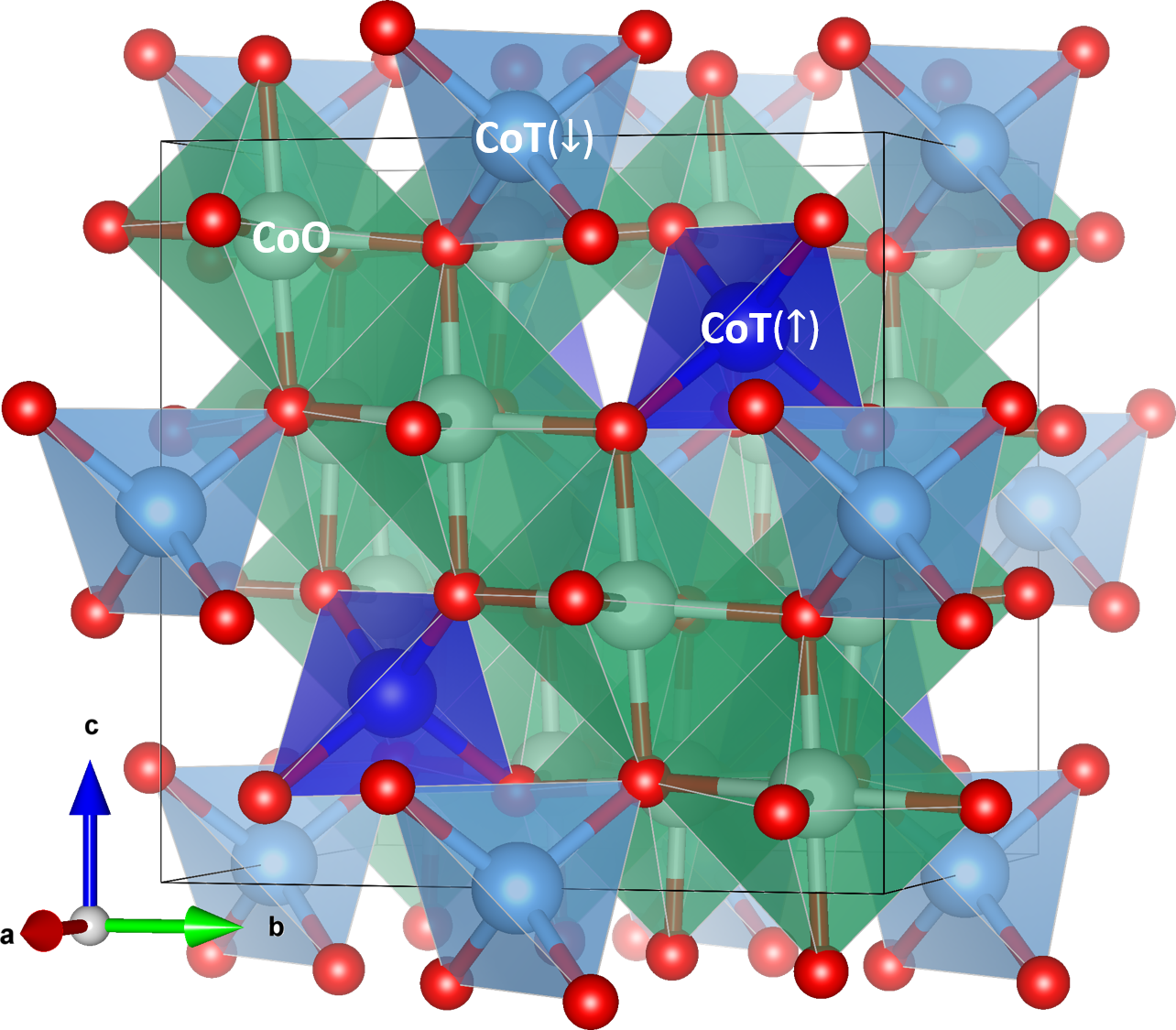}
\caption{Normal spinel atomic structure of \co. Octahedral Co are shown in green, tetrahedral Co are shown in blue/light blue (distinguishing spin polarization direction), and O are shown in red.}  \label{fig:struct}
\end{center}
\end{figure}


In this paper, we discuss the role of SPs (hereinafter referred to as ``polarons'' for simplicity) in tricobalt tetraoxide (\co), an anti-ferromagnetic oxide with a normal spinel structure. Despite that \co\ has been extensively investigated for a wide range of technologies \cite{wu2010,li2005,ma2014,wang2018,hamdani2010,aijaz2016,meher2011,xia2011}, a fundamental understanding of the optical properties of this material remains largely lacking, and conflicting results have been reported, e.g., for the band gap of bulk \co. For instance, a value of 1.5-1.7~eV has been commonly reported for the optical gap of \co~\cite{lohaus2016,jiang2014,waegele2014,shinde2006,belova1983,lima2014,cheng1998}. On the other hand, several experimental studies conclude that, despite a transition being observed around 1.5-1.7~eV, the true band gap of \co\ is significantly smaller, yielding a value of around 0.7-0.9~eV~\cite{qiao2013,singh2014,sousa2019,martens1985}. This conclusion, however, is not supported by time-resolved optical spectroscopy measurements, which suggest the state at $\sim$0.8~eV above the valence band maximum is a localized polaron state \cite{waegele2014,jiang2014,wheeler2012}. Along this direction, other experiments have indicated that the intrinsic carriers in \co\ are hole polarons that are characterized by a nearest-neighbor hopping conduction mechanism; and such a signature implies that hole polarons may affect the optical properties of \co\ in a similar way as in \ce{Fe2O3} \cite{ngamou2010,tronel2006,shinde2006,lohaus2016,sahoo2013,sparks2018,cheng1998,koumoto1981}. Collectively, the existing results indicate that much is left to be understood regarding the nature of the optical transitions near the band edge of \co, and how it is related to polaron formation.

The aim of this work is to resolve the conflicting results in the literature on \co, and provide a coherent description of its electronic structure, carrier conduction, and optical properties through first-principles calculations. In particular, we discuss the level of theory needed for a proper description of the electronic structure of the material. In addition, we elucidate the role of polaron formation on the electronic band gap and optical spectra of p-doped \co, and we discuss how uniaxial strain can be used to distinguish polaron related transitions in the optical spectra. This work provides a straightforward method for considering SP effects in the optical absorption, alongside unambiguous SP peak assignment in agreement with previous experimental studies of \co. Furthermore, our study distinguishes the role of SP formation from other optical factors that are often considered in the literature (e.g. exciton formation, thermal broadening of optical spectra and electron-phonon renormalization of band edges). Beyond \co\, our study presents a roadmap for first-principles calculations in the investigation of SP effects on the optical properties of TMOs.


\begin{figure}[t]
\begin{center}
\includegraphics[keepaspectratio=true,width=9cm]{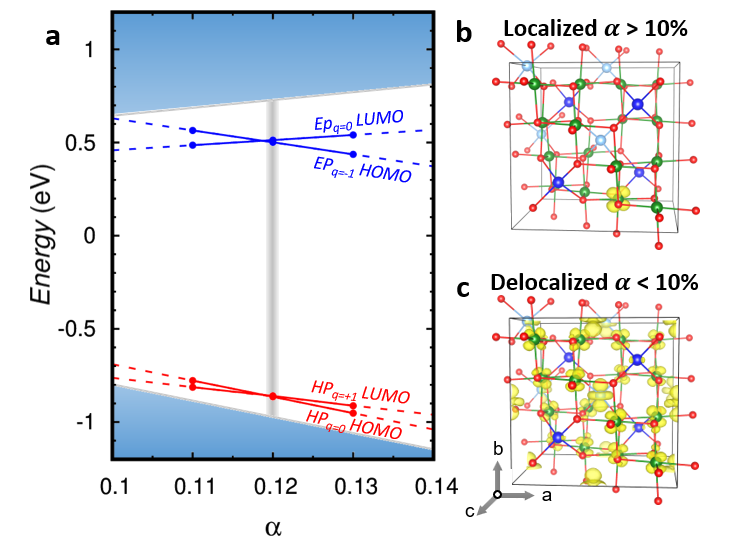}
\caption{\textbf{a.} Generalized Koopmans' condition for electron polaron (EP) and hole polaron (HP) in \co. The exact exchange $\alpha$ for the PBE0($\alpha$) method is varied until the condition $\text{HOMO}_q=\text{LUMO}_{q+1}$ (at fixed geometry where polaron has formed) is met. In both cases, we find that at an exact exchange of 0.12 Koopmans' condition is satisfied. The corresponding pristine gap is computed to be 1.70 eV. \textbf{b.} Localized and \textbf{c.} delocalized hole wavefunction, subject to the value of the exact exchange. Isosurface plots use a cutoff of 10\% the maximum.}
\label{fig:IPEA}
\end{center}
\end{figure}

We begin by discussing our computational strategy for addressing the electronic properties of \co{} (spinel structure shown in Figure \ref{fig:struct}). It is well known that density functional theory (DFT), with conventional local or semi-local exchange-correlation functionals, is not sufficient to provide a proper description of polarons in transition metal oxides and often severely underestimates the band gap of these materials \cite{dudarev1998}. In order to mitigate this issue, several approaches have been proposed, including DFT+$U$ with an orbital specific Hubbard $U$ correction, and hybrid functional that includes a fraction of Hartree-Fock exchange ($\alpha$), hereinafter denoted as PBE0($\alpha$). However, these calculations are known to highly depend on the choice of the Hartree-Fock exchange or Hubbard $U$ correction. For instance, a wide range between 0.8 and 2.0~eV has been reported in the literature for the band gap of \co, depending on the level of theory employed \cite{singh2014}. In this context, it is also necessary to emphasize that, despite significant development has been made, establishing a first-principles approach that allows for an accurate prediction of the electronic properties of TMOs remains a significant challenge \cite{kent2018toward}.

Here, we implement both hybrid functional and DFT+$U$ calculations to provide an unbiased description of the electronic properties of \co. Notably, in variation with previous hybrid functional calculations, we invoked the generalized Koopmans' condition to determine the value of $\alpha$ from first-principles. We stress that this strategy has been shown to successfully predict the band gap of materials with band gaps up to 14 eV and has been particularly successful for polaronic systems \cite{miceli2018,smart2018,liu2018electron,lany2011predicting}. Specifically, the generalized Koopmans' condition enforces the condition $\text{IP}_q=\text{EA}_{q+1}$ at a fixed geometry for an isolated state in the materials, where $\text{IP}_q$ is the ionization potential of an occupied state at the charge state $q$, whereas $\text{EA}_{q+1}$ is the electron affinity of the same state when it is unoccupied at the charge state $q+1$.  Here, we determine the value of $\alpha$ by enforcing the condition $\text{IP}_q=\text{EA}_{q+1}$ for both the hole and electron polaron (see Figure~S1 for more details of the electron polaron), and we obtained a value of 0.12 for the Hartree-Fock exchange $\alpha$ in both cases, as shown in Figure \ref{fig:IPEA}. We note that hole polarons do not form for $\alpha$ below 10\%, as illustrated in Figure \ref{fig:IPEA}b-c). The value of $\alpha$ determined in this manner is an intrinsic property of the bulk system, and the choice of defect used to enforce the Koopmans’ condition is proper as long as it has minimum hybridization with the bulk Bloch states \cite{bischoff2019adjustable,miceli2018,smart2018}.

We then determined $U$ parameters based on the hybrid functional results, and the stable formation of electron and hole polarons (see Table~S1 for more details). In particular, we find that the use of $U$ values of $U_\text{Co(O)}=4$ eV and $U_\text{Co(T)}=3$ eV provides consistent results with the Koopmans' compliant hybrid functional (band gap agrees within 0.1~eV), as well as a proper description of the electronic properties of the system, as discussed later in this communication. We also note that hole polarons do not form for $U$ values below 2.5 eV (see Figure~S2).

\begin{figure}[t]
\begin{center}
\includegraphics[keepaspectratio=true,width=8cm]{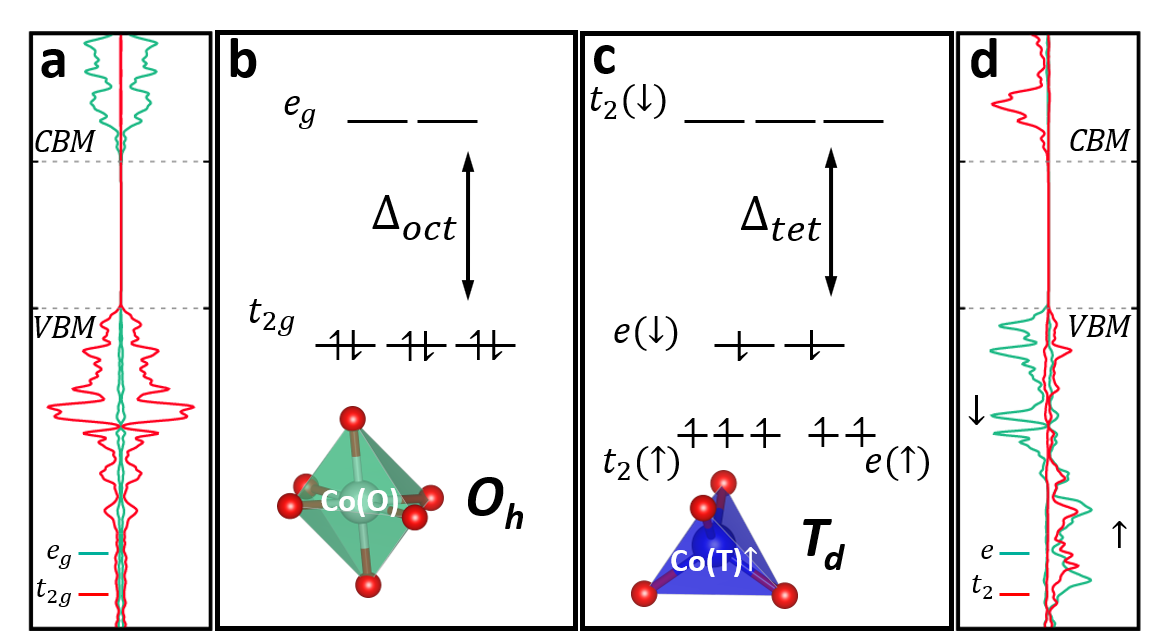}
\caption{\textbf{a.} Projected density of states (PDOS) of Co(O) on $t_{2g}$ and $e_g$ orbitals. Schematic representation of the electronic configuration of \textbf{b.} octahedral Co$^{3+}$ and \textbf{c.} tetrahedral Co$^{2+}$ in \co\ due to a crystal field splitting ($\Delta_{oct}$ and $\Delta_{tet}$, respectively). \textbf{d.} PDOS of Co(T) on $t_2$ and $e$ orbitals. All the PDOS was computed with DFT+$U$.}
\label{fig:split}
\end{center}
\end{figure}

All the calculations were then carried out using the plane-wave code Quantum ESPRESSO \cite{QE1} with norm-conserving pseudopotentials \cite{ONCV1}. A plane  wave cutoff of 100~Ry was used in all PBE+$U$ calculations, while a reduced cutoff of 50~Ry was implemented for the more demanding hybrid functional calculations (geometry and electronic structure are converged at 50 Ry). The calculations were generally performed within the 56 atom cubic cell with a $2\times 2\times 2$ k-point mesh for integration over the Brillouin zone. In addition, a $\sqrt{2}\times\sqrt{2}\times 2$ supercell with 224 atoms was utilized for comparison, particularly to understand the finite-size effects on polaron formation. Nevertheless, we find that the wavefunction character, energy level splitting, and band structure look largely unchanged between the 56 atom cubic cell and the 224 atom supercell (see Figure~S3 for further details). In all calculations with electron or hole polarons, the charged cell correction scheme as developed in Ref.~\citenum{PingJCP} was employed, which is necessary in order to remove the spurious interactions of the polarons with their periodic images and with the uniform compensating background charge \cite{kokott2018first}. For the rest of the manuscript, unless otherwise noted, the results presented here were obtained at the DFT+$U$ level of theory that is computationally less demanding compared to hybrid functional calculations.



To set a baseline for the discussion of polaron effects in \co, we briefly summarize the electronic structure of the pristine system. As shown in Figure \ref{fig:struct}, \co\ assembles in a normal spinel structure, where two thirds of Co occupy octahedral sites (denoted by Co(O)) and the remaining third occupy tetrahedral sites (denoted by Co(T)). We find that Co(O) exhibits a low-spin $3d^6$ orbital configuration with the $O_h$ symmetry, leading to filled $t_{2g}$ and empty $e_g$ bands, as shown in the calculated projected density of states (PDOS) and band diagram presented in Figure~\ref{fig:split} a-b. On the other hand, Co(T) with the $T_d$ symmetry forms a high-spin $3d^7$ configuration with a half-filled $e_g$ band, yielding an overall magnetic moment of $\sim$3.2 $\mu_B$ as already reported in experiments \cite{roth1964}. These Co(T) sites experience an anti-ferromagnetic interaction mediated through a super-exchange of mutually bonded oxygen; in addition, the presence of a large Hund's spin exchange results in a splitting of the $t_{2}$ band. This is shown in Figure \ref{fig:split} c-d, where we find that the $t_{2}$ minority spin states are formed at a higher energy level, whereas all majority spin states occur at lower and similar energies \cite{lima2014,chen2011} (also see Figure~S4 for further details). Overall, the behavior of the spin states and band splitting presented here are consistent with results reported in existing theoretical studies \cite{chen2011,wu2016}. 


Next, we discuss the nature of polaron formation in \co. Our calculations show that, among the Co(O) and Co(T) sites where hole polarons can form, the total energy of a hole polaron forming at Co(O) is lower than that of Co(T) by at least 70~meV. A more stable formation of the polaron on Co(O) is also reflected in the calculated density of states of \co\, where a larger contribution of Co(O) $d$ states is found at the valence band edge compared to the Co(T) $d$ states (see Figure~S4). Finally, our conclusion is consistent with the experimental study reported by Ngamou \textit{et al.} \cite{ngamou2010}, where the authors show that the hopping of polarons takes place in the octahedral sites and are responsible for driving the electrical transport in the oxide. Collectively, these results indicate that the computational approach employed here provides a proper description of polaron formation in \co.

\begin{figure*}
\begin{center}
\includegraphics[keepaspectratio=true,width=0.98\linewidth]{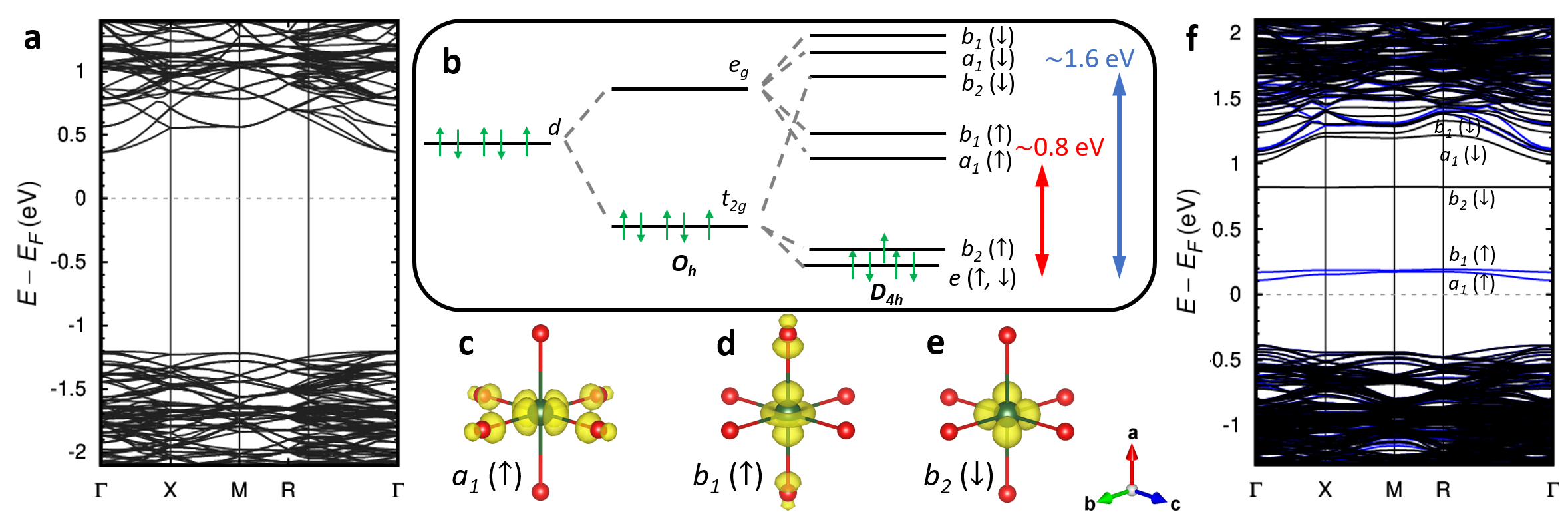}
\caption{\textbf{a.} Pristine band structure of \co\ with a 224 atom supercell (primitive cell band structure shown in SI Figure~S5). \textbf{b.} Hole polarons create a low-spin (LS) $d^5$ configuration at Co(O) along with a Jahn-Teller (JT) distortion which results in a $D_{4h}$ configuration and the creation of several mid-gap states. \textbf{c-e.} Wavefunction isosurface plots (yellow cloud) of the three polaron induced states under hole formation of $a_1(\uparrow)(d_{x^2-y^2})$, $b_1(\uparrow)(d_{z^2})$, and $b_2(\downarrow)(d_{xy})$ character, respectively. Isosurface plots use a cutoff value of 10\% the maximum. \textbf{f.} Band structure of \co\ with a hole polaron which shows several induced gap states (blue = spin up, black = spin down).}  \label{fig:hole}
\end{center}
\end{figure*}

Beyond the findings on the thermodynamical stability of hole polaron formation in \co, our calculations show that the hole polaron at Co(O) leads to several mid-gap states. As shown in Figure \ref{fig:hole}, we find that upon the hole polaron formation, Jahn-Teller (JT) distortion occurs at the Co(O) site due to an uneven occupation of the $t_{2g}$ band, and splits the degeneracy of the $O_h$ states. In addition, the uneven occupation of up/down states splits the spin degeneracy due to the on-site Coulomb repulsion of the $d$ orbitals. Specifically, we find that the majority spin states (e.g. $b_2(\uparrow), a_1(\uparrow), b_1(\uparrow)$) are located at a lower energy, whereas the minority spin states (e.g. $b_2(\downarrow), a_1(\downarrow), b_1(\downarrow)$) are pushed higher in the energy. Such splitting and ordering is consistent with the previous time-resolved spectroscopy measurements of Co(O) \cite{wu2016,belova1983}. 
In addition, these mid-gap states are consistent with the experiment reported in Ref.~\citenum{jiang2014}, where it was found that mid-gap excitations are associated with $a_1(\uparrow)$ and $b_1(\uparrow)$ states of Co(O). As a result, our analyses point to significant effects of hole polarons on the electronic structure of \co, most notably in the formation of mid-gap states in a similar way as found in \ce{Fe2O3}.


We now turn to a more quantitative discussion of polaron effects on the electronic structure of \co. In particular, we obtain a band gap of 1.6~eV and 1.7~eV with the current choice of $U$ and $\alpha$ (respectively), which is in excellent agreement with the value of 1.5-1.7~eV reported in Refs.~\citenum{lohaus2016,jiang2014,waegele2014,shinde2006,belova1983,lima2014,cheng1998}. However, this is significantly larger than the result of 0.7-0.9~eV claimed by other experiments. \cite{qiao2013,singh2014,sousa2019,martens1985}
We note that exciton binding energies are usually less than 150 meV in many TMOs \cite{ping20132,le2014,Ping2013}, and we show later that the calculated optical spectra, by including excitonic effects, cannot explain the low energy transition at 0.7-0.9~eV.
Interestingly, as shown in Figure~\ref{fig:hole}b, at the current level of theory, we find that the mid-gap states are located at 0.8~eV away from the valence band maximum, and are associated with the hole polaron formation. This observation suggests that a scenario similar to the one observed for \ce{Fe2O3} may also occur in \co, i.e., the true gap of \co\ is $\sim$1.6~eV, and the mid-gap states are responsible for the transitions found at $\sim$0.8 eV in the experimental optical absorption spectra \cite{qiao2013,singh2014,sousa2019,martens1985}.
In order to verify our hypothesis and to further elucidate the role of polaron formation, we calculated the optical absorption spectra of \co\ with and without a hole polaron, and we compared the results with available experimental data. Absorption spectrum with a hole polaron was computed with a 56-atom supercell \cite{concentrationNote}, which better represents experimental hole concentrations of p-type \co\ \cite{waegele2014,lohaus2016} due to abundant cation vacancies \cite{tronel2006,godillot2013}.

Therefore, we computed the imaginary part of the dielectric function in the random phase approximation (RPA) with local field effects (as shown in Figure 5) and solving the Bethe-Salpeter equation that includes excitonic effects (as shown in SI Figures S8 and S9), as implemented in the YAMBO-code \cite{YAMBO}, using the single particle eigenvalues and wavefunctions derived from DFT+$U$. We note that this choice of starting point considers the balance between accuracy and computational cost, similar to this work in Ref.~\citenum{Claudia2012}. In addition, for a direct comparison with experimental measurements, we calculated the absorption coefficient from the dielectric function \cite{jackson1999}.
\begin{align}
    A(\omega)=\frac{\omega}{c}\frac{\epsilon_2(\omega)}{\sqrt{\frac{\epsilon_1(\omega)+\sqrt{\epsilon_1(\omega)^2+\epsilon_2(\omega)^2}}{2}}}
\end{align}
\begin{figure}[t]
\begin{center}
\includegraphics[keepaspectratio=true,scale=1]{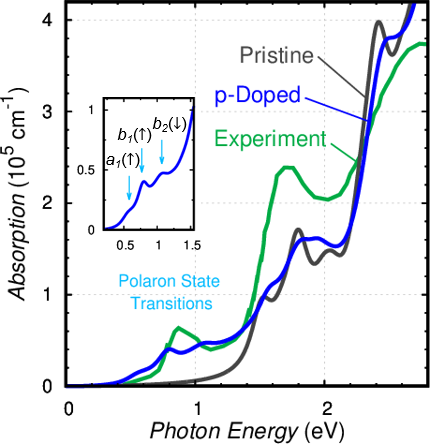}
\caption{Optical absorption of \co\ in the pristine system (black) and the p-doped system (blue). Notably, only p-type doping i.e. the formation of holes, will cause mid-gap transitions below 1.6 eV, in agreement with experimental optical spectrum of \co\ shown in green \cite{qiao2013}. The inset image displays the mid-gap transitions which are labeled according to the states formed from hole polaron formation as in Figure \ref{fig:hole}.
(Theoretical spectrum is an average of spectra with light polarized in the [100], [010], and [001] directions.)}  \label{fig:abs}
\end{center}
\end{figure}

The calculated optical absorption spectra of \co\ are shown in Figure~\ref{fig:abs}, together with the experimental spectrum. We find that the introduction of hole polarons leads to the formation of several lower energy optical transition peaks between 0.6 and 1.2 eV in \co. More importantly, we find that, in sharp contrast to the result obtained for the pristine system, the spectrum computed for \co\ with a hole polaron is in very good agreement with experimental data, where three lower lying transitions were also found between 0.7 and 1.1~eV \cite{qiao2013,lohaus2016,jiang2014,waegele2014}. Our analysis indicates that these transitions can be associated with those occur between the $p$-$d$ hybridized dispersive valence states and the localized $d$ states formed at the hole polaron site, for which the wavefunctions are illustrated in Figure \ref{fig:hole} c-e. Collectively, these results clearly support the interpretation that the true optical gap of \co\ is $\sim$1.6 eV and that the optical transitions observed at $\sim$0.8 eV are due to hole polaron formation at Co(O) sites.
We note that electron polarons do not lead to the formation of low lying transitions (see Figure~S6), indicative of the p-doped nature of the experimental system. 

In order to rule out the possibility that the low energy transitions $\sim$0.8 eV are caused by large excitonic effects~\cite{Laskowski2009,Bruneval2006,Wiktor2018,Rodl2012}, we also computed absorption spectra of pristine \co\ including excitonic effects by solving the Bethe-Salpeter Equation, as shown in SI Figure~S7, S8. More detailed discussions can be found in SI. 
Overall, no extra peaks in the BSE spectra show up at the energy range below 1 eV for pristine \co\ (no hole polaron included), which confirms the excitonic effects do not explain the low energy transitions in the absence of SPs. 

In addition, we note that we neglect electron-phonon coupling and thermal expansion effects on the band edge positions and absorption spectra at finite temperature, as discussed in Refs.~\citenum{monserrat2018phonon,bravic2019finite}. In the presence of temperature effects, the absorption edge may be subject to a red-shift, in addition to an overall broadening of the spectra, which is the subject of our future studies. However, we expect that these effects will not lead to an additional peak that are well separated from the main absorption in a direct band-gap semiconductor like \co, and therefore our conclusion on the contribution of small polarons to the low energy optical transitions still holds.


Finally, we propose an experimentally viable method for distinguishing optical transitions involving localized polaron states from  traditional  band-band  bulk  state  transitions. For \co, the JT distortion upon the introduction of the hole polaron extends the Co-O bonds along the $C_4$ axis, and this distortion may occur along any of the bond axis that aligns with the [100], [010], [001] directions of the cubic unit cell. Such a three-fold degeneracy can be broken if uniaxial strain is applied to the system along one of the crystal lattice directions, which in turn may affect the optical absorption spectrum.  In this regard, monitoring the change in the optical spectrum of \co\ in the presence of an uniaxial strain could potentially provide signatures of hole polarons associated with a specific JT distortion.

For demonstration, we considered a 1\% tensile strain applied along the [100] direction. We find that the three-fold degeneracy in the polaron states is broken upon the introduction of the strain; in particular, polaron formation with the JT elongation along the [100] direction is  lowered  in  the  energy  by  5  meV  compared  to  those  associated  with  the  [010]  or  [001] directions.   Here,  to  investigate  the  collective  and  individual  effects  of  these  polarons  on the absorption spectrum, we computed a thermally averaged ensemble spectrum by using a Boltzmann probability distribution of the optical absorption obtained for each case:
\begin{align}
    P_i = \frac{e^{E_i/kT}}{\sum_i e^{E_i/kT}} \quad , \quad A(\omega)=\sum_{i}P_i\,A_i(\omega)\label{eq:bolt},
\end{align}
where $E_{i}$ and $A_{i}$ are the energy and absorption spectrum of the system containing a polaron in the state $i$ ($i$ denotes different JT elongation direction), respectively. 

\begin{figure}
\begin{center}
\includegraphics[keepaspectratio=true,width=7cm]{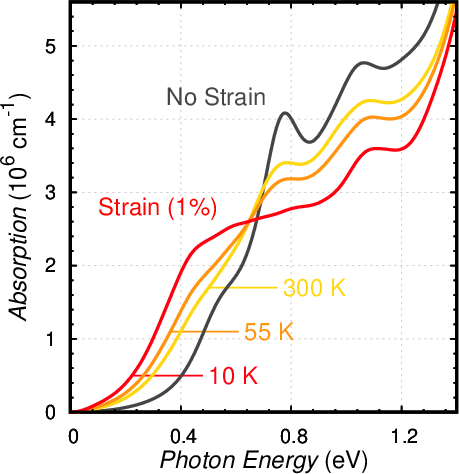}
\caption{Optical absorption of p-doped \co\ under 1\% uniaxial tensile strain along the [100] direction. Temperature dependence determines the probability for which direction the Jahn-Teller elongation will occur (as the degeneracy is removed under strain) and results in a red-shift of optical peaks related to the hole polaron.}  \label{fig:strain}
\end{center}
\end{figure}

The calculated optical absorption spectrum presented in Figure \ref{fig:strain} clearly shows a red-shift in the first peak that is associated with polarons. In particular, we find that at lower temperatures where $kT$ is on the order of 5 meV or less, the resulting optical spectra follow that of the lowest energy polaron associated with JT elongation along the [100] direction. At higher temperatures, the clear red-shift remains, although a high temperature of 300~K is sufficient to quench the 5~meV energy difference between polaron states. In contrast to the transition associated with hole polarons, we find that bulk band-band transitions at higher energy (above 1.5 eV) remain mostly unchanged upon uniaxial strain (see Figure~S9). Accordingly, this allows one to clearly distinguish the local polaron state involved in optical transitions, whose JT distortion renders them quite sensitive to strain, from that of the band-band bulk state transitions which are insensitive to strain.


To summarize, we present a detailed investigation of the electronic structure and polaronic induced optical transitions in \co\ based on first-principles calculations. We resolved several contradicting findings in the literature related to the character of the charge carrier and band gap of the material. In particular, we show that the optical gap of pristine \co\ is 1.6 eV, whereas the lower lying transition around $\sim$0.8~eV is associated with the hole polaron, which was misinterpreted as the band edge of the material. We also demonstrated the important effects of uniaxial strain on the optical spectra of \co, which in turn can be used to reveal the localized character of polaron-induced electronic states. 

Our study also suggests a strategy for establishing a potential first-principles approach that can simultaneously achieve an accurate description of polaron states, electronic band structure and optical properties in polaronic magnetic oxides. Specifically, the generalized Koopmans' condition can be utilized to derive the fraction of exact exchange from first-principles, which in turn can be used in hybrid functional for investigating the electronic structure of the oxide. These hybrid functionals can also be used for benchmarking DFT-$U$ calculations, which offer a much lower computational cost, or to provide inputs for higher level electronic structure methods, such as many-body perturbation theory within the \textit{GW} approximation. 

\medskip 

This work was performed under the auspices of the U.S. Department of Energy by Lawrence Livermore National Laboratory under Contract No. DE‐AC52‐07NA27344. T.S, T.A.P and T.O are supported by the U.S. Department of Energy, Office of Energy Efficiency and Renewable Energy, Fuel Cell Technologies Office. Y.P. is supported by the National Science Foundation under grant no. DMR-1760260 and CHE-1904547. This research used computational support from the LLNL Grand Challenge Program, the Center for Functional Nanomaterials, which is a US DOE Office of Science Facility, and the Scientific Data and Computing center, a component of the Computational Science Initiative, at Brookhaven National Laboratory under Contract No. DE-SC0012704, the National Energy Research Scientific Computing Center (NERSC) a U.S. Department of Energy Office of Science User Facility operated under Contract No. DE-AC02-05CH11231, the Extreme Science and Engineering Discovery Environment (XSEDE) which is supported by National Science Foundation Grant No. ACI-1548562 \cite{xsede}. We acknowledge fruitful discussion with Lin-Wang Wang and Di-Jia Liu.



\onecolumngrid


\section*{Supplementary material}

# Supplemental Information to "Optical Absorption Induced by Small Polaron Formation in Transition Metal Oxides - The Case of $Co_3O_4$"

Tyler J. Smart,[1,2] Tuan Anh Pham,[2] Yuan Ping,[3] and Tadashi Ogitsu[2]

[1]*Department of Physics, University of California Santa Cruz, Santa Cruz, CA, 95064, USA*
[2]*Quantum Simulations Group, Lawrence Livermore National Laboratory, Livermore CA, 94551, USA*
[3]*Department of Chemistry and Biochemistry, University of California Santa Cruz, Santa Cruz, CA, 95064, USA*

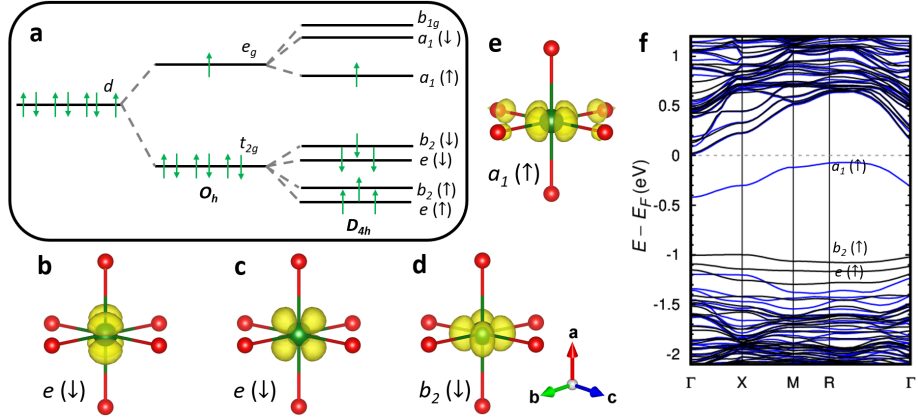

**Figure S1. a.** Electron polarons create a low-spin (LS) $d^7$ configuration at Co(O) along with a Jahn-Teller (JT) distortion which results in a $D_{4h}$ configuration and the creation of a few localized states, one of which is a mid-gap state, $a_1(\uparrow)$. **b-e.** Wavefunction isosurface plots (yellow cloud) of the four polaron induced states under electron polaron formation of $e(\downarrow)(d_{xz}$ and $d_{yz})$, $b_2(\downarrow)(d_{xy})$, and $a_1(\uparrow)(d_{x^2-y^2})$ character, respectively. Isosurface plots use a cutoff value of 10% the maximum. **f.** Band structure of $Co_3O_4$ with an electron polaron which shows induced gap states (blue = spin up, black = spin down).

**Hubbard $U$**

| $U_{Co(O)}$ | $U_{Co(T)}$ | Gap | HP forms? | EP forms? |
|---|---|---|---|---|
| 3 | 2 | 1.03 | yes | no |
| 3 | 3 | 1.38 | yes | no |
| 3 | 4 | 1.70 | yes | no |
| 3 | 5 | 1.53 | yes | no |
| 4 | 2 | 1.33 | no | yes |
| 4 | 3 | 1.57 | yes | yes |
| 4 | 4 | 1.71 | yes | yes |
| 4 | 5 | 1.80 | yes | yes |
| 5 | 2 | 1.41 | no | yes |
| 5 | 3 | 1.71 | no | yes |
| 5 | 4 | 1.89 | yes | yes |
| 5 | 5 | 2.01 | yes | yes |

**Table S1.** A $U$ on Co(O) must be equal to or greater than 4 eV in order to form the electron polaron (EP). For the hole polaron (HP), the $U$ on Co(O) can be as small as 3 eV. However, the value of $U$ on Co(T) must be similar to or

greater than that of Co(O). Namely if the $U$ on Co(T) is 2 eV smaller (or less) than the $U$ on Co(O), the hole polaron will not form. Lastly in order for the band gap to be reasonable, we select values of $U$ which agree with experimental values 1.5-1.7 eV. This leaves $U_{\text{Co(O)}} = 4$ eV and $U_{\text{Co(T)}} = 3 - 4$ eV as the only reliable choices for $U$, which also give consistent results with Koopman's compliant hybrid functional.

**Localization of Hole Polaron as a function of $U$**

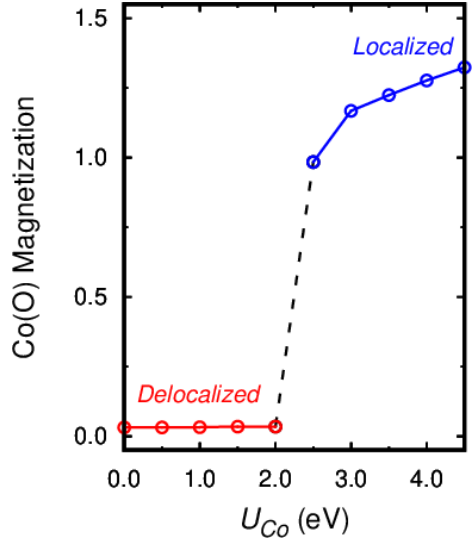

**Figure S2.** At Hubbard $U$ values greater than 2.5 eV holes localize on Co(O) with a magnetic moment $\gtrsim 1\,\mu_B$. Below $U = 2$ eV, holes delocalize in the system (charge is shared among the Co(O) in the cell so each have a slightly non-zero magnetic moment). For simplicity of discussions, here we set $U_{\text{Co(O)}} = U_{\text{Co(T)}} = U_{Co}$.

**Finite-Size Cell Convergence of Hole Polaron**

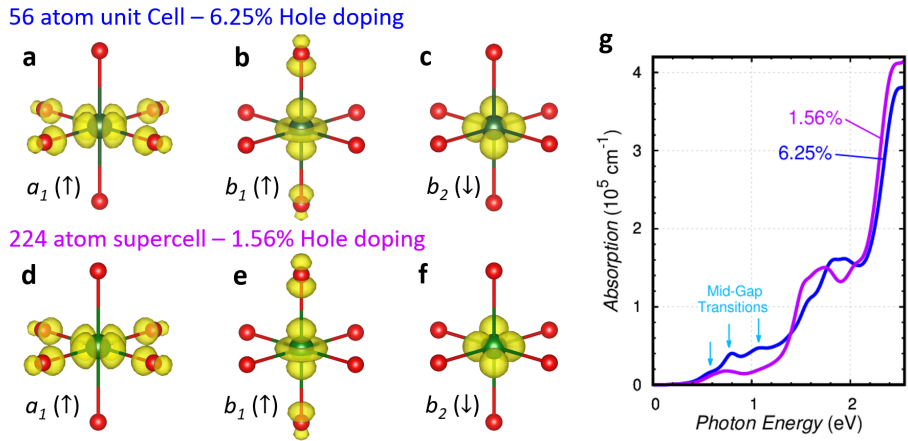

**Figure S3. a-c.** Hole polaron induced mid-gap wavefunctions in 56 atom calculation (12.5% concentration). **d-f.** Hole polaron induced mid-gap wavefunctions in a 224 atom supercell ($\sqrt{2} \times \sqrt{2} \times 2$) calculation which shows the same wavefunctions (3.125% concentration). **g.** Absorption spectrum of p-doped 224 atom supercell as compared with the 56 atom cell used in the main text. (Concentrations are defined as the percentage of holes per unit of Co3O4).

**Projected density of states of pristine $Co_3O_4$**

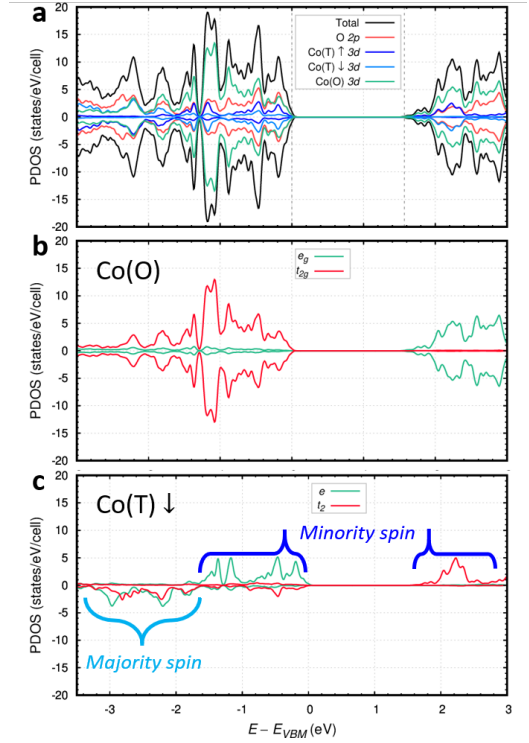

**Figure S4. a.** Density of states $Co_3O_4$ along with projected density of states on atomic orbitals. **b-c.** Projection of Co(O) and Co(T) $d$ states on $e_g/t_{2g}$ or $e/t_2$ orbitals showing splittings due to $O_h$ and $T_d$ symmetries (respectively). In the case of Co(T) in **c** further Hund's spin exchange results in lower energy majority spin states versus the minority spin states. In **c.** the density of states is multiplied by 2 to clarify the states.

**Comparison of Co(O) vs. Co(T) holes**

Conflicting results have been reported in the literature for the location of charge carriers in $Co_3O_4$. In particular, some studies mention that hole conduction occurs at tetrahedral sites [1–3], while others debate that conduction must occur at the octahedral sites [4–7]. From our DFT+$U$ calculations, we find that it is more likely for holes to form at Co(O) due to the larger density of states at the valence band edge corresponding to Co(O) $d$ compared to that of Co(T) $d$, as shown in Figure S5 and Figure S4. Furthermore, we compared the total energy of hole polaron formation at the Co(O) site versus the Co(T) and found that the total energy is always lower in the case of Co(O). For example with $U_{Co(O)} = 4$ eV and $U_{Co(T)} = 3$ eV, the total energy of the hole polaron at Co(O) is 70 meV lower than Co(T). We stress that this result is entirely lenient in the choice of $U$ and that we consistently see hole formation at Co(O) is easier. For example, with $U_{Co(O)} = U_{Co(T)} = 5$ eV, the total energy of Co(O) is lower by 280 meV than Co(T). In a final note, we find that only hole polarons at Co(O) can reproduce optical spectra of $Co_3O_4$ as in experiment (see Figure S6).

## DFT+U Band Structure and PDOS of $Co_3O_4$

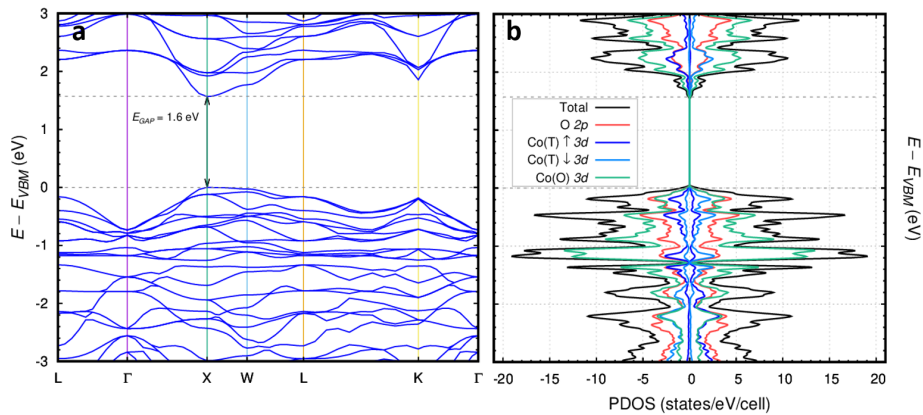

**Figure S5. a.** DFT+$U$ electronic band structure of $Co_3O_4$ (for the primitive cell) with highlighted the direct gap at X-X of 1.6 eV. **b.** Density of states $Co_3O_4$ along with projected density of states on atomic orbitals. The valence edge has major contribution from Co(O) 3$d$ and O 2$p$ with minor contribution of Co(T) 3$d$. The conduction edge has major contribution from Co(O) 3$d$ with minor contribution of Co(T) 3$d$ and O 2$p$.

## Other origins of mid-gap optical transitions

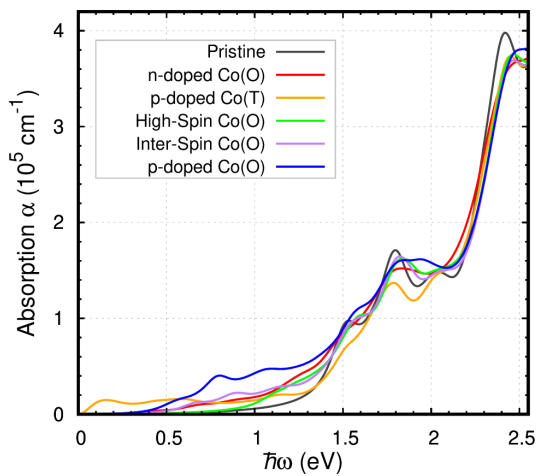

**Figure S6.** Optical absorption spectrum of $Co_3O_4$ with various potential sources of mid-gap state formation. Note that only p-doped $Co_3O_4$ with hole polarons formed at Co(O) yields mid-gap transitions in agreement with experiment. Further explanation of calculations: n-doped Co(O) = electron polaron formed on Co(O), p-doped Co(T) = hole polaron formed on Co(T), high-spin Co(O) = one Co(O) with high-spin formed (4 unpaired $d$ electrons), inter-spin Co(O) = one Co(O) with intermediate-spin formed (2 unpaired $d$ electrons), p-doped Co(O) = hole polaron formed on Co(O).

**Absorption Spectra of $Co_3O_4$ by solving the Bethe-Salpeter Equation**

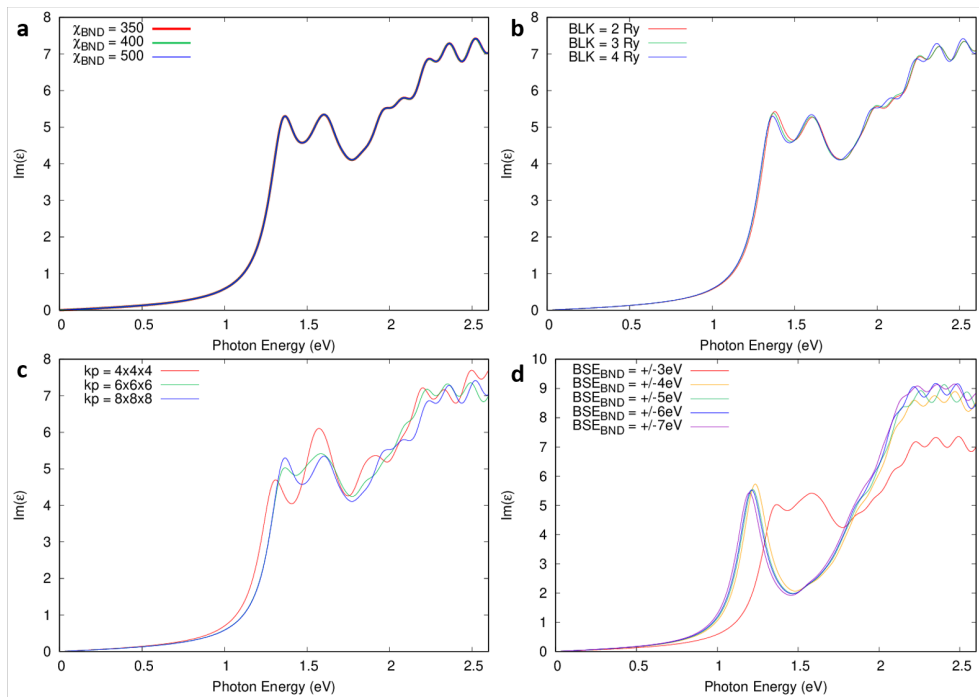

**Figure S7.** Convergence of BSE spectra with respect to **a.** bands used for calculating the dielectric matrices, **b.** BSE kernel and dielectric matrix block sizes, **c.** k-point sampling, and **d.** bands included in the BSE Hamiltonian.

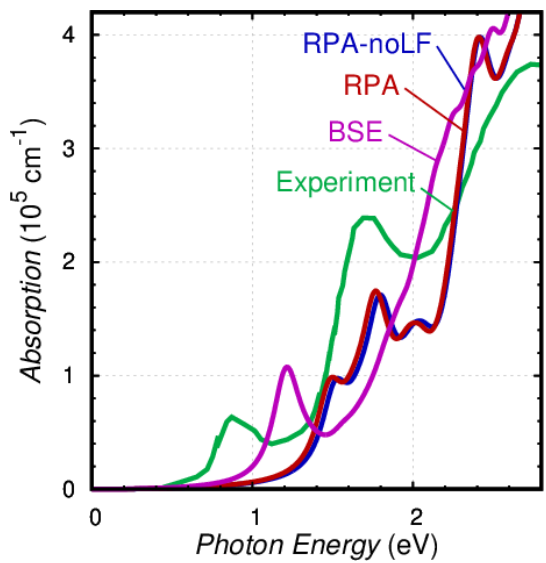

**Figure S8.** Comparison of theoretically predicted absorption spectra computed at RPA ("RPA-noLF" excludes local field effects and "RPA" includes local field effects) and BSE levels, alongside experimentally measured spectrum [8]. Both the RPA and BSE spectra are computed with the DFT+$U$ wavefunctions and eigenvalues as input as described in the main text.

We computed absorption spectra of pristine $Co_3O_4$ (without SPs) including excitonic effects by solving the Bethe-Salpeter Equation as shown in Figure S7 and Figure S8. The absorption edge is shifted to a lower energy by 0.3 eV compared to the one computed at the RPA level (with and without local field) as shown in Figure S8. This may indicate large exciton binding energy or strong excitonic effects in $Co_3O_4$ similar to several oxides discussed in past works [9–12]. However, this result needs to be considered with caution because it is well-known that the exciton binding energy can be strongly overestimated for transition metal oxides when lattice screening is not taken into account [13, 14].

**Effect of strain on the optical spectra of p-doped $Co_3O_4$**

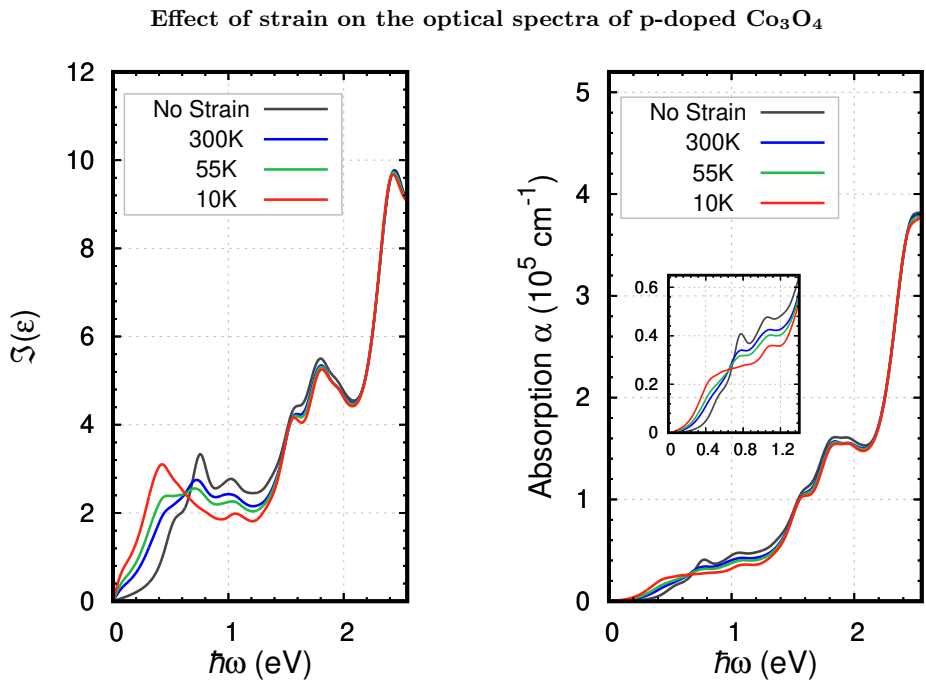

Figure S9. **a.** Imaginary part of the dielectric function and **b.** optical absorption under uniaxial strain of 1% increase along the (100) direction. Temperature dependence determines the probability in occupying polaron states which differ in JT orientation, which are non-degenerate upon uniaxial strain being applied into the system and yield the effects in the optical response observed. Note there is no clear shift in the bulk peak positions (above 1.5 eV), allowing for clear distinguishing of bulk vs. polaron related peaks.



\end{document}